\documentclass[12pt]{article}

\newenvironment{wileykeywords}{\textsf{Keywords:}\hspace{\stretch{1}}}{\hspace{
\stretch{1}}\rule{1ex}{1ex}}

\usepackage{amsmath,amssymb}
\usepackage{graphicx}
\usepackage{color}
\usepackage{dcolumn}
\usepackage{bm}
\usepackage[numbers,super,comma,sort&compress]{natbib}
\usepackage{pdflscape}

\newcommand{\onlinecite}[1]{\hspace{-1 ex} \nocite{#1}\citenum{#1}} 

\definecolor{background-color}{gray}{0.98}

\title{Efficient singular-value decomposition of the coupled-cluster triple 
excitation amplitudes}

\author{Michal Lesiuk\thanks{Faculty of Chemistry, University of Warsaw, 
Pasteura 1, 02-093 Warsaw, 
Poland, e-mail: lesiuk@tiger.chem.uw.edu.pl}}

\begin{document}

\maketitle

\begin{abstract}
We demonstrate a novel technique to obtain singular-value decomposition (SVD) of 
the coupled-cluster triple excitations amplitudes, $t_{ijk}^{abc}$. The presented method is based on 
the Golub-Kahan bidiagonalization strategy and does not require $t_{ijk}^{abc}$ to be stored. 
The computational cost of the method is comparable to several CCSD iterations. Moreover, the number 
of 
singular vectors to be found can be predetermined by the user and only those singular vectors which 
correspond to the largest singular values are obtained at convergence. We show how the subspace of 
the most important singular vectors obtained from an approximate triple amplitudes tensor can be 
used to solve equations of the CC3 method. The new method is tested for a set of small and 
medium-sized molecular systems in basis sets ranging in quality from double- to quintuple-zeta. It 
is 
found that to reach the chemical accuracy ($\approx 1$ kJ/mol) in the total CC3 energies as little 
as $5-15\%$ of SVD vectors are required. This corresponds to the compression of the $t_{ijk}^{abc}$ 
amplitudes by a 
factor of ca. $0.0001-0.005$. Significant savings are obtained also in calculation of interaction 
energies or rotational 
barriers, as well as in bond-breaking processes.
\end{abstract}

\begin{wileykeywords}
coupled-cluster theory, singular-value decomposition, electronic structure
\end{wileykeywords}

\clearpage

\begin{figure}[h]
\colorbox{background-color}{
\fbox{
\begin{minipage}{1.0\textwidth}
\begin{center}
\includegraphics[scale=1.50]{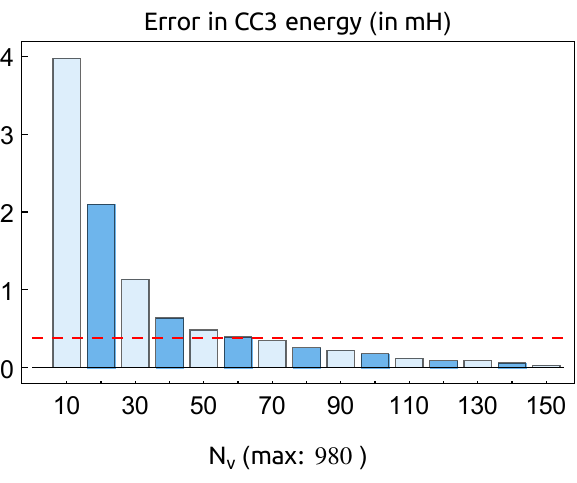} 
\end{center}
Error in the CC3 correlation energy (in mH) for water molecule (cc-pV5Z basis 
set) as a function of number of singular vectors ($N_v$) included in the 
expansion of triple excitation amplitudes. The horizontal red dashed line marks 
the $1$ kJ/mol accuracy threshold (the chemical accuracy).
\end{minipage}
}}
\end{figure}

\makeatletter
\renewcommand\@biblabel[1]{#1.}
\makeatother

\bibliographystyle{apsrev}

\renewcommand{\baselinestretch}{1.5}
\normalsize
\clearpage

\section*{Introduction}

Over the past several decades the coupled cluster (CC) theory 
\cite{coester58,coester60,cizek66,cizek69,cizek71,cizek72} has established 
itself as one of the most successful quantum chemical methods (see Ref. 
\onlinecite{bartlett07} for an extended survey). In particular, the CCSD(T) 
method of Raghavachari \emph{et al.} 
\cite{ragha89} serves as the \emph{gold standard} of 
quantum chemistry 
\cite{ragha96,scuseria90,tew07,rezac13a,rezac13b}. In well-behaved systems 
CCSD(T) is able to consistently deliver results of chemical accuracy (in the 
complete basis set limit) 
and thus provide a benchmark for other theoretical models. Some cases when the 
CCSD(T) method breaks down have also 
been intensively studied\cite{chan58,laidig87,piecuch02,piecuch04} and various 
extensions and corrections have been proposed. This includes the so-called 
externally 
corrected CC methods 
\cite{piecuch93b,stolarczyk94,piecuch96,li97,li98,li99,li00}, renormalized and 
completely renormalized CC approaches of Piecuch and collaborators 
\cite{kowalski00a,kowalski00b,kowalski01a,piecuch01a,piecuch01b,mcguire02,
kowalski01b,kowalski01c,
kowalski02,pimienta02,wloch06,piecuch05,piecuch04}, the CC(P;Q) 
hierarchy of techniques introduced recently
\cite{shen12a,shen12b,shen12c,shen12d,bauman17,magoulas18}, approaches based on 
partitioning of a similarity-transformed Hamiltonian 
\cite{gwaltney00a,gwaltney00b,gwaltney01,gwaltney02,hirata01,hirata04}, the 
orbital-optimized methods \cite{sherill98,krylov00,nooijen00,bozkaya12}, and 
techniques which incorporate the so-called $\Lambda$ amplitudes 
\cite{kucharski98b,musial10,taube08a,taube08b}.

An important observation related to the higher-order clusters ($T_n$) in the CC 
theory is that only a relatively small number of excitations (or 
linear combinations thereof) contribute significantly to the correlation energy 
and other results. It is therefore natural to attempt to extract the important 
information from $T_n$, \emph{i.e.} reduce the rank of a high-rank tensor 
$T_n$, and the singular-value decomposition (SVD) \cite{trefethen97} is a 
prominent method for accomplishing such task. It has been successfully used in 
numerous applications from signal processing \cite{lathauwer04,lathauwer07} and 
data analysis 
\cite{alter00,alter04} to psychometrics \cite{eckart36}. 

In the context of the CC theory the use of SVD has been first considered by 
Hino \emph{at al.} \cite{hino03,hino04} who have shown how CCSDT-1a 
equations can be solved in a SVD subspace. However, no efficient methods have 
been known thus far to compute SVD of a 
given $t_{ijk}^{abc}$ tensor.  In Ref. \onlinecite{hino04} this was achieved by 
diagonalization of a certain pseudo-density matrix. Unfortunately, the cost of 
computation of this 
matrix scales as $\mathcal{N}^8$ with the size of the system, $\mathcal{N}$, 
which is formally the 
same (up to a prefactor) as the cost of the complete CCSDT computations. This 
eliminates all 
potential gains from the compression of the $T_3$ amplitudes. Nonetheless, the 
results of Ref. 
\onlinecite{hino04} are very important as they suggest that the overall idea is 
sound provided that 
a more economical method of computing SVD of $t_{ijk}^{abc}$ tensor can be 
found. The main purpose 
of this work is to establish such a method.

We have adopted a set of important requirements which must simultaneously be 
fulfilled for the 
method to be of any use in quantum chemistry. First, the $T_3$ amplitudes need 
not to be stored in 
memory; some portion of the $t_{ijk}^{abc}$ tensor can be stored, \emph{e.g.} 
after some 
pre-screening is applied, but this should be an option, not a requirement. 
Second, the 
computational 
cost of the procedure should be smaller than $\mathcal{N}^8$, for the reasons 
explained earlier. 
Third, the method should be able to selectively find only the most important 
singular vectors, i.e. 
those that correspond to the largest singular values, and the desired number of 
vectors should be 
set by the user beforehand. Finally, the method should be as close to a 
black-box as possible and 
require a minimal user input and intervention.

To test the new method we employ a subspace formed by the most important 
singular vectors (obtained 
from SVD of an approximate $T_3$ amplitudes) to solve equations of iterative CC 
methods. It would 
be 
natural to apply this idea to the most complete CCSDT model. However, this 
requires careful 
consideration of the details of the implementation including factorization of 
several high-order 
terms and is beyond the scope of the present work. Instead, we concentrate on 
the CC3 method 
\cite{koch97} which is a successful and widely used approximation to the full 
CCSDT theory.

\section*{Theory}

\subsection*{Preliminaries}

Coupled cluster theory is based on exponential parametrization of the 
many-electron wavefunction
\begin{align}
\label{cc1}
 |\Psi\rangle = e^T\,|\phi_0\rangle,
\end{align}
where $|\phi_0\rangle$ is the reference determinant and $T=\sum_{n=1} T_n$ are 
the cluster operators expressed through the creation $a^\dagger, 
b^\dagger,\ldots$, and annihilation operators $i,j,\ldots$ as
\begin{align}
\label{cluster}
 &T_1 = \sum_{ai} t_i^a \,a^\dagger i, \\
 &T_2 = \frac{1}{4}\sum_{abji} t_{ij}^{ab} \,a^\dagger b^\dagger j i, \\
 &T_3 = \frac{1}{36}\sum_{abckji} t_{ijk}^{abc} \,a^\dagger b^\dagger c^\dagger 
k j i,
\end{align}
and so on. The indices $i, j, k, \ldots$ and $a, b, c, \ldots$ denote the 
occupied and virtual 
spin-orbitals, respectively. When the occupation of an orbital is not specified 
we use general 
indices $p, q, r, \ldots$. Throughout the paper the canonical Hartree-Fock 
determinant is assumed 
as 
the reference wavefunction and the spin-orbital energies are denoted by 
$\epsilon_p$. For further 
use, we also introduce a shorthand notation $\langle X\rangle 
\stackrel{\mbox{\tiny def}}{=} 
\langle 
\phi_0 | X \phi_0 \rangle$ and $\langle X|Y\rangle \stackrel{\mbox{\tiny 
def}}{=} \langle X \phi_0 
| 
Y \phi_0 \rangle$ for arbitrary operators $X$, $Y$. The electronic 
Schr\"{o}dinger Hamiltonian is 
divided into two parts, $H=F+W$, where $F$ is the Fock operator and $W$ is the 
fluctuation 
potential.

The conventional CC equations are obtained by inserting the Ansatz (\ref{cc1}) 
into the electronic 
Schr\"{o}dinger equation and projecting onto the manifold of singly, doubly, 
\emph{etc.}, excited 
states determinants. The electronic energy is given by the expression $E = 
\langle e^{-T} H e^T 
\rangle$. By truncating the cluster operator at a certain level one obtains 
approximate CC models. 
For example, by setting $T=T_1+T_2$ and projecting onto $\langle \,_{i}^{a} |$, 
$\langle 
\,_{ij}^{ab} |$ one recovers the standard CCSD theory 
\cite{purvis82,scuseria87,stanton91,hampel92}. 
An analogous truncation $T=T_1+T_2+T_3$ combined with projection onto $\langle 
\,_{i}^{a} |$, 
$\langle \,_{ij}^{ab} |$, $\langle \,_{ijk}^{abc} |$ gives the CCSDT theory 
\cite{noga87,scuseria88}. 

A popular method to reduce the computational burden connected with the inclusion 
of the $T_3$ 
operator is to invoke a perturbative approach. For example, in the CCSD[T] 
method 
\cite{bartlett90,watts93} the amplitudes $T_3$ are approximated by the 
leading-order perturbative 
expression
\begin{align}
\label{t32}
 T_3^{[2]} = (\epsilon_{ijk}^{abc})^{-1} \langle \,_{ijk}^{abc} | \big[ W, T_2 
\big]\rangle \,a^\dagger b^\dagger c^\dagger 
k j i,
\end{align}
where 
$\epsilon_{ijk}^{abc}
=\epsilon_i+\epsilon_j+\epsilon_k-\epsilon_a-\epsilon_b-\epsilon_c$ is 
the three-particle energy denominator. This leads to a relatively simple energy 
correction which is 
added on top of CCSD
\begin{align}
\label{e4t}
E^{[4]}_T = \langle T_2 | \big[ W, T_3^{[2]} \big] \rangle. 
\end{align}
Working expressions of the CCSD(T) method are virtually the same apart from one 
term, $\langle \,T_1 | \big[ W, T_3^{[2]} \big]\rangle$, which involves 
singly excited configurations \cite{ragha89}. The computational cost of 
evaluating $E^{[4]}_T$ 
scales as $\mathcal{N}^7$.

A different (non-perturbative) way of simplifying the CCSDT method relies on 
adopting 
approximations 
to the $T_3$ amplitudes equations yet retaining the iterative nature of the 
theory. Several 
variants 
such as CCSDT-n, $n=1,2,3$, of Urban \emph{et al.} \cite{urban85,noga87b} were 
proposed but in the 
present paper we rely on the CC3 method introduced by Koch \emph{et al.} 
\cite{koch97} The singles 
and doubles equations in the CC3 method are exactly the same as in CCSDT but the 
triples equation 
is 
simplified to the form
\begin{align}
\label{r3cc3a}
 \langle \,_{ijk}^{abc} | \big[ F, T_3 \big] + \big[ \widetilde{W}, T_2 \big] 
\rangle = 0,
\end{align}
where $\widetilde{W} = e^{-T_1} W e^{T_1}$. The CC3 method scales as 
$\mathcal{N}^7$ which is 
formally the same as CCSD(T). In practice, CC3 is considerably more expensive 
than CCSD(T) due to 
its iterative nature but still orders of magnitude cheaper than the full CCSDT.

For convenience of the readers let us briefly recall the most important 
properties of the 
singular-value decomposition. An in-depth discussion of this topic can be found, for example, in 
Ref. \onlinecite{trefethen97}. SVD is a factorization of an arbitrary $m\times n$ 
rectangular matrix 
${\bf M}$ to the form
\begin{align}
\label{svd1}
 {\bf M = U\,\Sigma V^\dagger },
\end{align}
where the following statements are valid about the matrices $\bf U$, 
$\bf\Sigma$, and $\bf V$
\begin{itemize}
\item ${\bf U}$ is an $m \times m$ unitary matrix collecting 
orthonormal eigenvectors of ${\bf MM^\dagger}$ (left-singular vectors);
\item ${\bf V}$ is an $n\times n$ unitary matrix collecting orthonormal 
eigenvectors of ${\bf 
M^\dagger M}$ (right-singular vectors);
\item ${\bf \Sigma}$ is a rectangular $m\times n$ diagonal matrix with 
\emph{non-negative} real 
numbers on the diagonal (singular values).
\end{itemize}
The singular values are identical to square roots of non-zero eigenvalues of 
${\bf MM^\dagger}$ (or 
${\bf M^\dagger M}$). One of the most useful properties of SVD is that the best 
(in the sense of 
the 
square norm) rank-$r$ approximation to a matrix ${\bf M}$ can be obtained by 
retaining on the diagonal of $\bf\Sigma$ the largest $r$ singular values and 
neglecting the rest.

\subsection*{Decomposition of the $T_3$ amplitudes}

Throughout the paper, we treat the triple amplitudes tensor $t_{ijk}^{abc}$ as a 
three-dimensional 
tensor with collective indices $ai$, $bj$, and $ck$. Thus, the dimension of a 
tensor is $OV$, where 
$O$ is the number of occupied orbitals and $V$ is the number of virtual 
orbitals. Moreover, the 
tensor is symmetric with respect to an exchange of all three indices. The 
ordering of the orbitals 
$a$, $i$ within the collective index $ai$ is irrelevant as long it is used 
consistently in all 
expressions.

Unfortunately, in three dimensions there exists no decomposition which retains 
all the merits of 
the 
two-dimensional SVD like the optimal truncation property. Therefore, many 
decomposition strategies 
have been proposed which possess some desirable properties. In the context of 
quantum chemistry, 
the 
canonical product decomposition\cite{carroll70,benedikt11,benedikt13,bohm16} or 
tensor 
hypercontraction decomposition 
\cite{hohenstein12a,hohenstein12b,hohenstein12c,hohenstein13,parrish14} serve as 
prime examples. 
However, the decomposition (or compression) of the $T_3$ amplitudes tensor 
employed here relies on 
the so-called Tucker-3 format \cite{tucker66} which in the present case reads
\begin{align}
\label{tuck1}
&t_{ijk}^{abc} \approx \sum_{XYZ}^{N_{v}} t_{XYZ} \,U^X_{ai} 
\,U^Y_{bj} \,U^Z_{ck},
\end{align}
or equivalently
\begin{align}
\label{tuck2}
&\hat{T}_3 \approx \sum_{XYZ}^{N_{v}} t_{XYZ}\,\hat{U}^X \hat{U}^Y 
\hat{U}^Z,\;\;
&\hat{U}^X = \sum_{ai}U^X_{ai}\, a^\dagger i.
\end{align}
One can say that $t_{XYZ}$ is a compressed triple amplitudes tensor in the 
subspace spanned by all 
possible combinations of $\hat{U}^X$. Let us introduce the \emph{compression 
factor} 
$\rho\in[0,1]$, 
defined through the relation $N_v=\rho OV$, which measures how successful the 
compression is. Clearly, in the limit $N_v \rightarrow OV$ (or 
$\rho\rightarrow1$) the decomposition (\ref{tuck2}) 
becomes exact. Note that the same decomposition as in Eq. (\ref{tuck1}) has 
been employed by Hino \emph{at al.} \cite{hino03,hino04} to solve equations of 
the CCSDT-1a method.

The advantage of the Tucker format is that it comes with a prescription on how 
to select the 
optimal 
$U^X_{ai}$ (see Refs. \onlinecite{lathauwer00,grasedyck10,vannie12} for an 
extended discussion). 
First, one performs ``flattening'' of the $t_{ijk}^{abc}$ tensor, \emph{i.e.} 
rewrites it as a 
two-dimensional $O^2V^2\times OV$ matrix, $t_{aibj,ck}$. Next,  SVD of the 
``flattened`` matrix is 
performed. The right-singular vectors form the desired tensors $U^X_{ai}$ whilst 
the left-singular 
vectors can be discarded. The optimal truncation is achieved by selecting those 
tensors $U^X_{ai}$ 
that correspond to the largest singular values of the ``flattened`` matrix.

\subsection*{Compressed CC3 method}

The main idea of the compressed CC methods is to perform CC iterations with the 
$T_3$ cluster 
operator given in the form (\ref{tuck2}). The matrices $U^X_{ai}$ required to 
form the expansion in 
Eq. (\ref{tuck2}) are obtained by performing SVD of some approximate $T_3$ 
amplitudes which must be 
known in advance. They can be obtained by carrying out the CCSD calculations 
first and evaluating 
$T_3^{[2]}$. This is the choice adopted in this work but many other options are 
possible, 
\emph{e.g.} taking $T_3$ from MRCI calculations within some active-orbital 
space.

A complete algorithm for computing SVD of an arbitrary $t_{ijk}^{abc}$ tensor is 
presented in 
Section III. Here we assume that the necessary matrices $U^X_{ai}$ are known and 
discuss an optimal 
implementation of the compressed CC3 theory. As a by-product of the SVD 
procedure the matrices 
$U^X_{ai}$ obey the orthonormality condition
\begin{align}
\label{uai1}
 \sum_{ai} U^X_{ai}\, U^Y_{ai} = \delta_{XY}.
\end{align}
As argued in Ref. \onlinecite{hino04} significant simplifications can be 
achieved if one performs 
an 
orthogonal rotation of the matrices $U^X_{ai}$ so that the following relation is 
fulfilled
\begin{align}
\label{uai2}
\sum_{ai} U^X_{ai}\, U^Y_{ai}\, (\epsilon_i-\epsilon_a) = \epsilon_X 
\delta_{XY},
\end{align}
where $\epsilon_X$ are some real-valued constants. This rotation preserves the 
orthonormality 
condition (\ref{uai1}) and is lossless in terms of the information carried by 
$U^X_{ai}$.

In the CC3 theory the $T_3$ amplitudes are given by Eq. (\ref{r3cc3a}) which can 
be rewritten to a 
more explicit form
\begin{align}
\label{r3cc3b}
\epsilon_{ijk}^{abc}\,t_{ijk}^{abc} + \langle \,_{ijk}^{abc} | \big[ 
\widetilde{W}, T_2 \big] 
\rangle = 0. 
\end{align}
Upon inserting the compressed form of the amplitudes, Eq. (\ref{tuck1}), and 
making use of Eqs. 
(\ref{uai1}) and (\ref{uai2}) one arrives at
\begin{align}
\label{comptxyz}
 \left( \epsilon_X + \epsilon_Y + \epsilon_Z \right) t_{XYZ} = \sum_{abcijk} 
U^X_{ai} 
\,U^Y_{bj} \,U^Z_{ck} \langle \,_{ijk}^{abc} | \big[ \widetilde{W}, T_2 \big] 
\rangle,
\end{align}
where $\epsilon_X$ have been defined through Eq. (\ref{uai2}).
Explicit expression for the matrix element on the right-hand-side of the above 
formula is given in 
Ref. \onlinecite{koch97}. In the closed-shell case it reads
\begin{align}
\label{rxyz}
 \langle \,_{ijk}^{abc} | \big[ \widetilde{W}, T_2 \big] \rangle = 
 P_{ijk}^{abc} \Big[ \sum_d t_{ij}^{ad} (ck\widetilde{|}bd) - \sum_l 
t_{il}^{ab} 
(ck\widetilde{|}lj) 
\Big],
\end{align}
where $P_{ijk}^{abc}$ is a permutation operator
\begin{align}
\label{pijk}
 P_{ijk}^{abc} = \left(\,_{ijk}^{abc}\right) + \left(\,_{ikj}^{acb}\right) + 
\left(\,_{jik}^{bac}\right) + \left(\,_{kij}^{cab}\right) + 
\left(\,_{jki}^{bca}\right) + 
\left(\,_{kji}^{cba}\right),
\end{align}
and $(pq\widetilde{|}rs)$ denotes the dressed ($T_1$ similarity-transformed) 
two-electron integrals 
\cite{koch97}. Note that without the compression the computational cost of 
evaluating Eq. 
(\ref{rxyz}) scales as $O^3V^4$ in the leading-order term. This typically 
constitutes a bottleneck 
in the conventional CC3 calculations. In order to evaluate the expression on the 
right-hand-side of 
Eq. (\ref{comptxyz}) we define a handful of intermediate quantities
\begin{align}
\label{ijk1}
 &I^X_{ai} = \sum_{bj} t_{ij}^{ab}\,U^X_{bj}, \\
 \label{ijk2}
 &J^X_{ab} = \sum_{ci} (ab\widetilde{|}ic)\,U^X_{ci}, \\
 \label{ijk3}
 &K^X_{ij} = \sum_{ka} (ij\widetilde{|}ka)\,U^X_{ak}.
\end{align}
The computational cost of calculating $I^X_{ai}$, $J^X_{ab}$, and $K^X_{ij}$ 
tensors scales as $N_v 
O^2 V^2 = \rho\,O^3 V^3$, $N_v O V^3 = \rho\,O^2 V^4$, and $N_v O^3 V = 
\rho\,O^4 V^2$. In 
practice, 
their evaluation is implemented as a series of matrix-matrix multiplications 
using BLAS routines. 
We 
have never found this step to be particularly time-consuming, even for large 
values of $\rho$.

Expressing the right-hand-side of Eq. (\ref{comptxyz}) in terms of the 
intermediates 
(\ref{ijk1})$-$(\ref{ijk3}) yields the final relation for the compressed 
amplitudes $t_{XYZ}$
\begin{align}
\label{txyz}
 \left( \epsilon_X + \epsilon_Y + \epsilon_Z \right) t_{XYZ} = P_{XYZ} \Big[ 
I_{ai}^X\, U_{bi}^Y\, 
J_{ba}^Z - I_{ai}^X\, U_{aj}^Y\, K_{ij}^Z \Big],
\end{align}
where $P_{XYZ}$ is a permutation operator analogous to Eq. (\ref{pijk}) but 
involving a sum over 
all 
possible permutations of the indices $X$, $Y$, $Z$. Evaluation of the first term 
in the square 
brackets in Eq. (\ref{txyz}) limits the efficiency of the algorithm. The second 
term in Eq. 
(\ref{txyz}) is by a factor of $O/V$ less expensive. For an efficient 
implementation of Eq. 
(\ref{txyz}) it is critical that the multiplications of the intermediate 
matrices are performed in 
two steps, \emph{e.g.} $I_{ai}^X \left( U_{bi}^Y\,J_{ba}^ Z \right)$. In this 
particular case the 
first step scales as $\rho^2\, O^3 V^4$ whilst the second as $\rho^3 O^4 V^4$. 
We found the former 
step to be the most time-consuming in almost all test calculations reported 
further in the text. 
Taking into account that the evaluation of the uncompressed CC3 triple 
amplitudes, Eq. 
(\ref{r3cc3b}), scales as $O^3 V^4$ in the leading-order term, we can conclude 
that the compression 
reduces this effort by a factor of $\rho^2$.

It may be disconcerting that the computational cost of the second step in Eq. 
(\ref{txyz}) scales as $\rho^3 O^4 V^4$. This would be equivalent to 
$\mathcal{N}^8$ if $O$ and $V$ were 
simultaneously increased, and the value of $\rho$ were kept fixed. However, as 
shown numerically in the next sections, when $O$ and $V$ are simultaneously 
doubled, the optimal 
$\rho$ is approximately halved. As a result, the asymptotic cost of evaluating 
Eq. (20) under these circumstances is proportional to $\mathcal{N}^5$. One can 
also consider a different scenario where the values of $\rho$ and $O$ are kept 
fixed and only $V$ is increased. This corresponds to a situation where, e.g. a 
small basis set is used to determine $\rho$ and then this $\rho$ is 
subsequently employed in calculations for the same system with larger basis 
sets. In this case the computational cost of evaluating Eq. (20) scales as 
$\mathcal{N}^4$, formally the same as the most expensive term in CCSD (with 
$O$ fixed).

With the compressed triple amplitudes calculated from Eq. (\ref{txyz}), the 
remaining task in the CC3 iteration cycle is to compute  $T_3$ contributions to 
the $T_1$ and $T_2$ amplitude equations, see Eqs. (100) and (101) in Ref. 
\onlinecite{koch97}. In our pilot implementation this is accomplished by 
reconstructing (on-the-fly) the triples amplitudes tensor by using Eq. 
(\ref{tuck1}). Next, it is inserted into Eqs. (100) and (101) in Ref. 
\onlinecite{koch97} to give the final result. Despite this approach is not 
optimal it has never been found to be the limiting factor in the calculations 
reported in this work.

\section*{Efficient SVD of $T_3$ amplitudes}

\subsection*{Golub-Kahan bidiagonalization}

In this section we present an efficient iterative algorithm to find a predefined 
number of singular 
vectors of a given $t_{ijk}^{abc}$ amplitudes tensor. We begin by recalling the 
key expressions of 
the Golub-Kahan bidiagonalization method \cite{golub65} which forms a backbone 
of the present SVD algorithm.
Any $m \times n$ rectangular matrix $\mathbf{A}$ can be brought into the 
bidiagonal form
\begin{align}
 \bf A = P B Q^\dagger,
\end{align}
where $\bf P$ and $\bf Q$ are $m\times m$ and $n\times n$ unitary matrices, and 
$\bf B$ is a 
$m\times n$ upper bidiagonal matrix. The proof of this theorem is constructive 
and based on the 
following double recursive formulas
\begin{align}
\label{gk1}
 &\alpha_j p_j = {\bf A} q_j - \beta_{j-1} p_{j-1},\\
 \label{gk2}
 &\beta_j q_{j+1} = {\bf A}^\dagger p_j - \alpha_j q_j,
\end{align}
where $p_j$ and $q_j$ are the $j$-th columns of $\bf P$ and $\bf Q$, 
respectively, and the 
constants $\alpha_j$ and $\beta_j$ are chosen so that $p_j$ and $q_j$ are 
normalized. It can be shown that the bidiagonal matrix assumes the following 
form
\begin{align}
{\bf B} = \left[
\begin{matrix}
  \alpha_1 & \beta_1 \\
   & \alpha_2 & \beta_2 & &  \\
   &  & \alpha_3 & \beta_3  \\
   &  &  & \ddots & \ddots \\
   &  &  &  & \alpha_{n-1} & \beta_{n-1} \\
   &  &  &  &  & \alpha_n
 \end{matrix}
\right]
\end{align}
and can be padded with zeros to match the required rank. There are two major 
advantages of the 
Golub-Kahan procedure. First, it is straightforward to calculate SVD of the 
matrix $\bf B$ and then 
reconstruct SVD of the full matrix $\bf A$.
Indeed, both $\bf B B^\dagger$ and $\bf B^\dagger B$ are square tridiagonal 
matrices and several 
robust techniques are available in the literature to diagonalize a tridiagonal 
matrix (see Ref. 
\onlinecite{coakley13} and references therein). Second, the recursive formulas 
(\ref{gk1}) and 
(\ref{gk2}) contain only products of the matrices $\bf A$ and $\bf A^\dagger$ 
with the vectors 
$q_j$ 
and $p_j$, respectively. They can be calculated on-the-fly without explicit 
storage of the full 
$\bf 
A$ matrix. Unfortunately, in order to obtain accurate singular vectors with the 
help of Eqs. 
(\ref{gk1}) and (\ref{gk2}) one would need to perform a complete 
bidiagonalization. This is both 
expensive and wasteful since only a small percentage of the dominant singular 
vectors is typically 
needed in practice.

\subsection*{Iterative restarted SVD}

Assume one wants to find $r$ singular vectors of the matrix ${\bf A}$ which 
correspond to the 
largest singular values. One first selects some dimension $k$ of the search 
space which must be somewhat larger 
than $r$. In the first step (initiation) $k$ steps of the Golub-Kahan 
bidiagonalization are 
performed. In the matrix notation we may write
\begin{align}
\label{gk3}
 &{\bf A} {\bf Q}_k = {\bf P}_k {\bf B}_k, \\
\label{gk4}
 &{\bf A}^\dagger {\bf P}_k = {\bf Q}_k {\bf B}_k^\dagger + \beta_k q_{k+1} 
e_k^\dagger,
\end{align}
where ${\bf P}_k$ and ${\bf Q}_k$ are rectangular matrices composed of the first 
$k$ columns of 
$\bf 
P$ and $\bf Q$, respectively, ${\bf B}_k$ is  the leading $k\times k$ principal 
sub-matrix of $\bf 
B$, and $e_k$ is a vector of dimension $k$ with unity in the last position. The 
second term on the 
right-hand-side of Eq. (\ref{gk4}) can be viewed as a remainder which vanishes 
in the limit of the 
complete bidiagonalization. In the case of a partial bidiagonalization it allows 
us to continue the 
process (restart) by simply employing $q_{k+1}$ in the next step, Eq. 
(\ref{gk1}). 

After $k$ steps of the initial bidiagonalization one computes SVD of the ${\bf 
B}_k$ matrix, i.e. , 
${\bf B}_k = {\bf X}_k\,{\bf \Sigma}_k {\bf Y}_k^\dagger $. By inserting this 
formula back into 
Eqs. 
(\ref{gk3}) and (\ref{gk4}) and rearranging we get
\begin{align}
\label{gk5}
&{\bf A} {\bf \bar{Q}}_k = {\bf \bar{P}}_k {\bf\Sigma}_k, \\
\label{gk6}
&{\bf A}^\dagger {\bf\bar{P}}_k = {\bf\bar{Q}}_k {\bf\Sigma}_k^\dagger + 
\beta_k\, q_{k+1}\, 
e_k^\dagger {\bf X}_k,
\end{align}
where ${\bf\bar{Q}}_k={\bf Q}_k {\bf Y}_k$ and ${\bf\bar{P}}_k={\bf P}_k {\bf 
X}_k$. Now we can 
select the largest $r$ singular 
values from ${\bf \Sigma}_k$ and shrink the decomposition back to $r$, i.e. 
temporarily reduce the 
search space size to $r$ (collapse). This is done by simply sorting the diagonal 
elements 
($\Sigma_k$) in the descending order and neglecting all elements $k>r$ together 
with the 
corresponding vectors ${\bf\bar{Q}}_k$ and ${\bf\bar{P}}_k$. The resulting 
decomposition is given 
formally by Eqs. (\ref{gk5}) and (\ref{gk6}), but with $k$ replaced by $r$ in 
all instances.

The next phase of the procedure consists of performing additional steps of the 
Golub-Kahan 
bidiagonalization in order to increase the search size of the space back to $k$ 
and thus improve 
the 
quality of the desired $r$ singular vectors. Unfortunately,
the specific form of the remainder has been destroyed in Eq. (\ref{gk6}), so 
that it is no longer 
proportional to $q_{k+1}\,e_k^\dagger$ for some vector $q_{k+1}$. As argued 
earlier, this form must 
be restored in order to facilitate the search space expansion. For this purpose 
we adopt the method 
put forward by Baglama and Reichel \cite{baglama05}. The idea is to expand the 
search space by one 
pair of vectors $r\rightarrow r+1$ which are purposefully chosen so that the 
correct bidiagonal 
form 
analogous to Eqs. (\ref{gk3}) and (\ref{gk4}) is restored. This is accomplished 
by setting
\begin{align}
 {\bf \bar{Q}}_{r+1} = \big[ \,\bar{q}_1, \bar{q}_2, \ldots, \bar{q}_p, q_{k+1} 
\big],
\end{align}
where $q_{k+1}$ is taken from Eq. (\ref{gk4}), and
\begin{align}
 {\bf \bar{P}}_{r+1} = \big[ \,\bar{p}_1, \bar{p}_2, \ldots, \bar{p}_p, 
\bar{p}_{r+1} \big],
\end{align}
where $\bar{p}_{r+1}$ is obtained by normalizing the following vector
\begin{align}
 x = A q_{r+1} - \sum_{i=1}^r \rho_i \bar{p}_i,
\end{align}
where $\rho_i$ are numerical constants chosen to make $\bar{p}_{r+1}$ orthogonal 
to all previous 
trial vectors, $\bar{p}_i$, $i=1,\ldots,r$. This leads us to the following 
partial decomposition
\begin{align}
\label{gk7}
&{\bf A} {\bf \bar{Q}}_{r+1} = {\bf \bar{P}}_{r+1} {\bf B}_{r+1}, \\
\label{gk8}
&{\bf A}^\dagger {\bf\bar{P}}_{r+1} = {\bf\bar{Q}}_{r+1} {\bf B}_{r+1}^\dagger 
+ 
\bar{\beta}_{r+1}\, 
q_{k+2}\, e_{r+1}^\dagger,
\end{align}
which is suitable for restart. A minor inconvenience connected with the 
procedure of Ref. 
\onlinecite{baglama05} is that the matrix ${\bf B}_{r+1}$ is no longer 
bidiagonal. Indeed, it 
possesses the following structure
\begin{align}
\label{br1}
{\bf B}_{r+1} = \left[
\begin{matrix}
  \sigma_1 & & & \rho_1 \\
   &  \ddots & & \vdots \\
   &  & \sigma_r & \rho_r \\
   &  &  &  \bar{\alpha}_{r+1} \\
 \end{matrix}
\right],
\end{align}
where $\bar{\alpha}_{r+1}=||x||$. After restarting the Golub-Kahan 
bidiagonalization and expanding 
the search space back to size $k$ we obtain
\begin{align}
\label{bknew}
{\bf B}_{k} = \left[
\begin{matrix}
  \sigma_1 & & & \rho_1 \\
   &  \ddots & & \vdots \\
   &  & \sigma_r & \rho_r \\
   &  &  & \bar{\alpha}_{r+1} & \beta_{r+1} \\
   &  &  &  & \ddots & \ddots \\
   &  &  &  &  & \alpha_{k-1} & \beta_{k-1} \\
   &  &  &  &  &  & \alpha_{k} \\
 \end{matrix}
\right].
\end{align}
One can see that the above matrix has a ''spike`` composed of the constants 
$\rho_i$ which prevents 
it from achieving a proper bidiagonal form. This appears somewhat sub-optimal 
because one has to 
use 
general SVD algorithms to decompose ${\bf B}_{k}$  
which are typically less effective than dedicated procedures designed with 
bidiagonal matrices in 
mind. In practice, however, the matrix ${\bf B}_{k}$ is rather small and we have 
never found this 
step to be particularly troublesome.

To summarize, the iterative restarted SVD procedure described here consists of 
three phases. In the 
first phase (initiation) one performs $k$ steps of the ordinary Golub-Kahan 
bidiagonalization 
achieving the decomposition given by Eqs. (\ref{gk3}) and (\ref{gk4}). In the 
second phase 
(collapse) one computes SVD of the small matrix ${\bf B}_k$. This yields 
approximate singular 
values 
and vectors which are then sorted, and only the most significant vectors are 
retained in the search 
space reducing its size to $r$. In the third phase (expansion) of the procedure 
a pair of vectors 
is 
added to the search space, enabling to restart the Golub-Kahan procedure, see 
Eqs. 
(\ref{gk5})$-$(\ref{br1}). Finally, the size of the search space is increased to 
$k$ by performing 
a 
certain number of bidiagonalization steps. This allows to return to phase two 
and establishes an 
iterative procedure which is repeated until the desired number of singular 
values/vectors have 
converged to the prescribed tolerance.

\subsection*{Implementation and technical details}

Thus far we have not touched upon a very important aspect of the presented 
algorithm - selection of 
the initial vector $q_0$ for the bidiagonalization (guess). In principle, any 
non-zero vector can 
be 
used to initiate this process, see Eqs. (\ref{gk1}) and (\ref{gk2}). However, if 
the starting 
vector 
is unsuitably chosen the algorithm tends to give 
$\beta_k=0$ in Eq. (\ref{gk2}) after some number of iterations. This prevents 
the bidiagonalization 
process from progressing in the standard fashion and a new vector must be 
provided in order to 
continue. We have implemented and tested several possible sources of the 
starting vectors:
\begin{itemize}
 \item $T_1$ amplitudes;
 \item successive eigenvectors of the $T_2$ amplitudes tensor;
 \item random vector - elements are generated quasi-randomly on the interval 
$[-\alpha 
t_{ai},+\alpha t_{ai}]$, where $t_{ai}$ are the respective elements of the 
single amplitudes 
tensor, 
and $\alpha$ is a positive constant.
\end{itemize}
The first guess is probably the most efficient in terms of number of iterations 
but has a 
considerable disadvantage - once $\beta_k=0$ is encountered at some step of the 
bidiagonalization 
there is no natural way to continue the process.
The $T_2$ guess performs only marginally poorer but offers a straightforward way 
to restart after 
$\beta_k=0$ - 
one simply selects the next eigenvector in the order of increasing eigenvalues. 
The random vector 
guess is a reasonable choice and has the advantage of being cheap and having an 
infinite supply of 
vectors for restart. Unfortunately, it also typically requires a larger number 
of iteration cycles 
to converge to a good accuracy.
To sum up, we found that the guess based on the eigenvectors of the $T_2$ 
amplitudes tensor is the 
best overall and the overhead connected with the diagonalization is manageable.

Another technical problem related to the bidiagonalization process, Eqs. 
(\ref{gk1}) and 
(\ref{gk2}), is the loss of orthogonality amongst the vectors $p_j$ and/or 
$q_j$. This occurs 
solely 
due to finite precision of the arithmetic. The simplest remedy to this problem 
is to perform full 
orthogonalization at each step, \emph{i.e.} the new vectors $p_n$ and $q_n$ are 
orthogonalized to 
all previous vectors $p_i$ and $q_i$, $i=1,\ldots,n-1$, respectively, by using 
the Gram-Schmidt 
procedure. The main drawback of this procedure is its high cost which grows as 
the iterations 
proceed (note that the size of the vectors $p_n$ is $O^2V^2$). A less expensive 
alternative to full 
orthogonalization has been proposed by Simon and Zha \cite{simon00} who observed 
that loss of 
orthogonality is mitigated to a sufficient extent when orthogonalization is 
performed only among 
$q_n$. This one-sided orthogonalization variant has been adopted in all 
calculations reported in 
this 
work.

Finally, let us analyze the computational costs and storage requirements of the 
presented 
algorithm. 
A single step of the Golub-Kahan bidiagonalization requires to multiply the 
matrix ${\bf A}$ 
separately from the left and from the right by some trial vectors. In the case 
of the tensor 
$t_{ijk}^{abc}=t_{aibj,ck}$ the computational cost of this operation 
asymptotically scales as 
$2\cdot O^3 V^3$. The initiation phase of the described algorithm requires to 
perform $k$ 
bidiagonalization steps and costs $2k\cdot O^3 V^3$. To estimate the remaining 
workload let us 
assume that after $n_{\mbox{\scriptsize it}}$ expansion-collapse cycles the 
desired $p$ singular 
vectors have converged to the desired precision. Each cycle consists of 
expanding the search space 
from $p+1$ to $k$ and thus requires $k-p-1$ bidiagonalization steps. Therefore, 
the total cost of 
the procedure is $2k\cdot O^3 V^3 + 2n_{\mbox{\scriptsize it}}(k-p-1)\cdot O^3 
V^3$. In other words, if the search space size $k$ is equal to the dimension of 
the flattened matrix, $OV$, the procedure described above is equivalent to the 
approach adopted by Hino \emph{et al.} and would possess $\mathcal{N}^8$ 
scaling of the computational costs.

The cost analysis from the previous paragraph assumes that the tensor 
$t_{ijk}^{abc}$ is stored. In 
our implementation only non-negligible elements of $t_{ijk}^{abc}$ are stored on 
the disk in a 
format 
which allows to read one row/column at the time. We have also developed a fully 
direct version of 
the algorithm where the elements of $t_{ijk}^{abc}$ are calculated on-the-fly as 
needed. The cost 
of 
the on-the-fly algorithm are obviously larger - the overhead depends strongly on 
the system size 
and 
on the efficiency of the prescreening, but it typically amounts to a factor of 
$5-10$.

One can see that the performance of the algorithm depends on a careful choice 
of 
$n_{\mbox{\scriptsize it}}$ and $k$ for a given $p$. In fact, larger $k$ 
typically require a 
smaller 
value of $n_{\mbox{\scriptsize it}}$, but increases the cost of the initiation 
phase. We found that 
the optimal choice of the search space size is $k\approx p + 10-20$. The only 
exception from this 
rule occurs where a large number of vectors is requested (half of the SVD total 
space size or 
more). 
In such cases the value of $k$ must be increased somewhat.

The theory presented here, along with the uncompressed CC3 method, was 
implemented in a locally 
modified version of the \textsc{Gamess} program package \cite{gamess1}. The 
validity of the new 
implementation was verified by comparing with independent routines available in 
the \textsc{Psi4} 
program package \cite{psi4}. On-the-fly and semi-direct variants of the 
Golub-Kahan 
bidiagonalization scheme, as described earlier in the text, are available. The 
current version of 
the code is fine-tuned for decomposition of the $T_3^{[2]}$ amplitudes, see Eq. 
(\ref{t32}), but it 
can easily be adapted for other sources of an approximate $t_{ijk}^{abc}$ 
tensor. The present 
implementation is restricted to closed-shell systems.

\section*{Numerical examples}

\subsection*{Total correlation energies}

In order to investigate the performance of the compressed CC3 method for 
computation of the 
correlation energies we performed calculations for 15 small and medium-sized 
molecules composed of 
the first- and second-row atoms. The geometries of the molecules considered here 
were taken from 
the 
G2$-$1 and G2$-$2 neutral test sets of Curtiss \emph{et al.} \cite{curtiss97} 
available on the 
World 
Wide Web \cite{g02}. For all diatomic and three-atomic molecules we used 
Dunning-type cc-pV$X$Z 
basis sets \cite{dunning89,dunning94} ranging in quality from $X=2$ to $X=5$. 
For larger molecules 
cc-pV$X$Z basis sets up to $X=4$ were employed. In all calculations reported 
here the threshold for 
convergence of the SVD vectors was $10^{-6}$ (in the square norm) and the size 
of the search space 
was fixed as $N_v+10$, where $N_v$ is the desired number of vectors to be found. 
Under these 
conditions the convergence of the iteration procedure was typically achieved in 
under 20 iterations 
and in many cases only $3-5$ cycles were sufficient. Spherical representation of 
the Gaussian basis 
set is employed in all calculations reported here. 

For each molecule in the test set we performed SVD-CC3 calculations with varying 
SVD subspace 
space, 
$N_v$ in Eq. (\ref{tuck1}). The value of $N_v$ was systematically increased (in 
steps of $10$) and 
the error in the correlation energy with respect to the exact CC3 result was 
recorded. In Table 
\ref{tab1} we show the compression levels $\rho$ and the number of SVD vectors 
which allow to reach 
the chemical accuracy ($1\,\mbox{kJ/mol} \approx 0.4\,\mbox{mH}$) of the total 
correlation energy.

Since the maximum possible size of the SVD subspace ($O\cdot V$) is different 
for each basis 
set/molecule the quantity $N_v$ is not transferable between systems. However, we 
claim that for a 
fixed basis set the value of $\rho$ should be (to some extent) transferable 
between systems of a similar size and thus can be used to 
estimate how many SVD vectors must be included to meet the adopted accuracy 
criteria. To confirm 
this the mean value of $\rho$ for each basis set and the corresponding standard 
deviation are 
reported in Table \ref{tab1}. One can see that $\rho=15\%$ is sufficient to 
reach the chemical 
accuracy in all systems under consideration and $\rho=10\%$ is adequate for a 
majority of them. 
According to the analysis undertaken in the previous section $\rho=15\%$ allows 
to compress the 
$T_3$ amplitudes tensor by a factor of $\approx 0.003$ and reduce the cost of 
evaluating Eq. 
(\ref{r3cc3b}) by a factor of $\approx 0.02$.

As presented in Table \ref{tab1} the average value of $\rho$ decreases 
significantly when passing 
from cc-pVDZ to cc-pVTZ. For larger basis sets the average $\rho$ is fairly 
constant. This 
observation is of practical interest as it allows to estimate the optimal $\rho$ 
for a given system 
from calculations in a small basis set. Moreover, it suggests that SVD-CC 
methods are expected to 
be 
equally useful in calculations with small and large basis sets as the efficiency 
of the compression 
is not significantly affected by changing the $V/O$ ratio.

\begin{figure*}
\begin{tabular}{ccc}
 \includegraphics[scale=0.90]{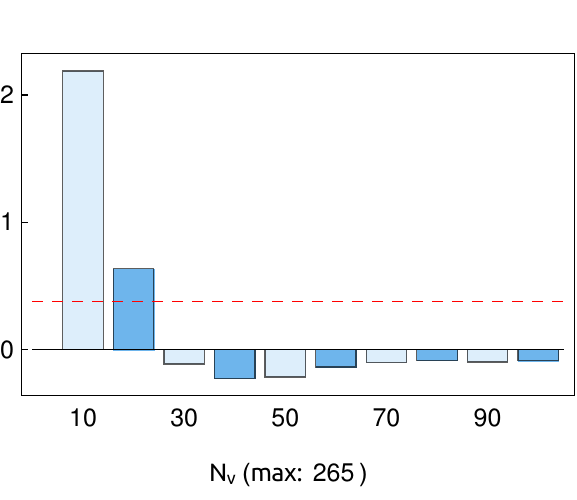} & 
 \includegraphics[scale=0.90]{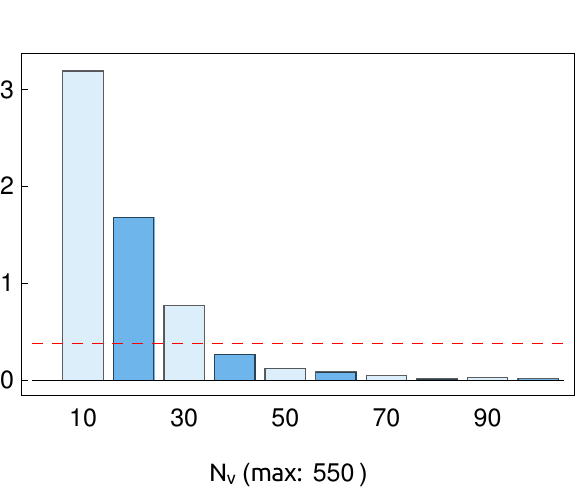} & 
 \includegraphics[scale=0.90]{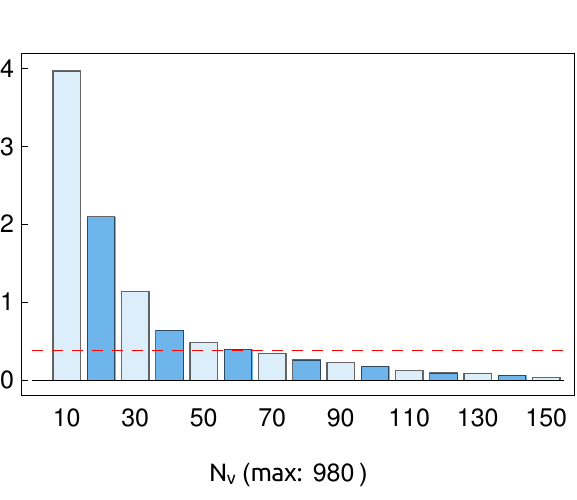} \\
 \includegraphics[scale=0.90]{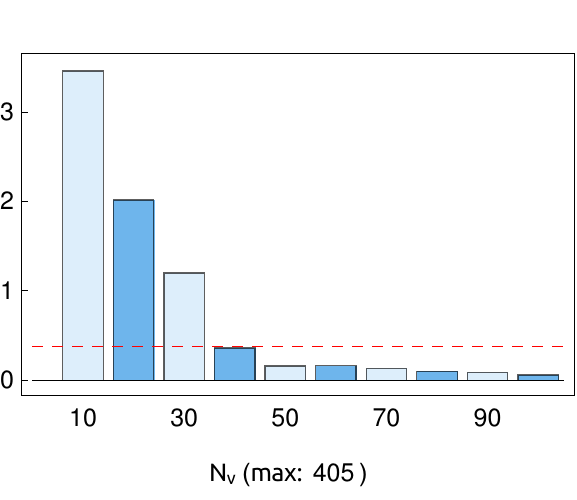} & 
 \includegraphics[scale=0.90]{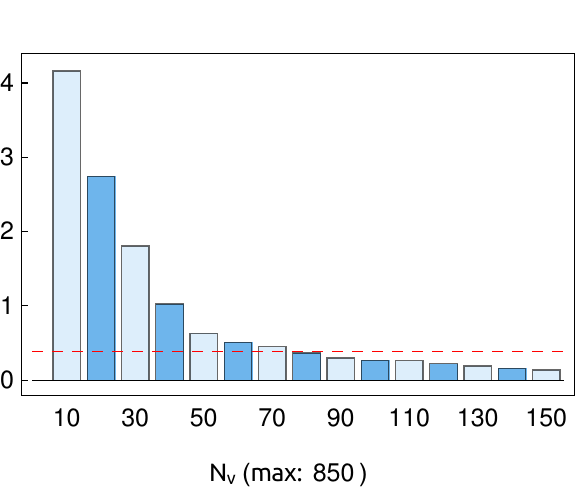} & 
 \includegraphics[scale=0.90]{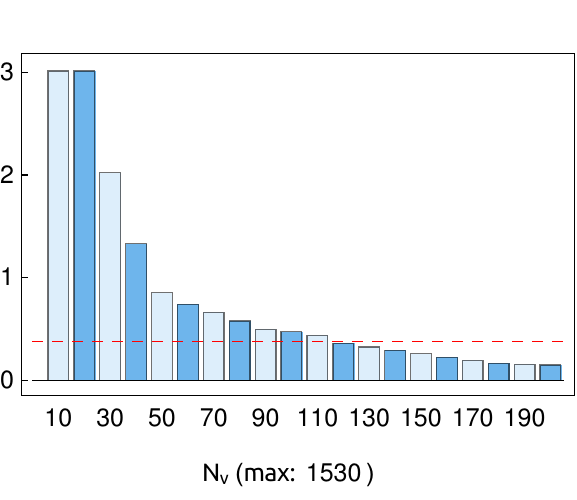} \\
 \includegraphics[scale=0.90]{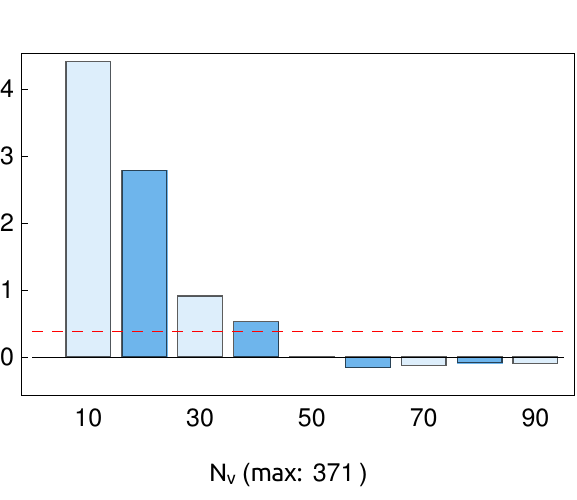} & 
 \includegraphics[scale=0.90]{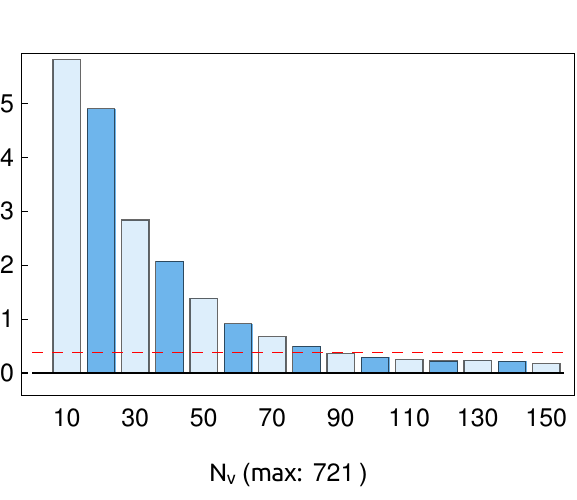} & 
 \includegraphics[scale=0.90]{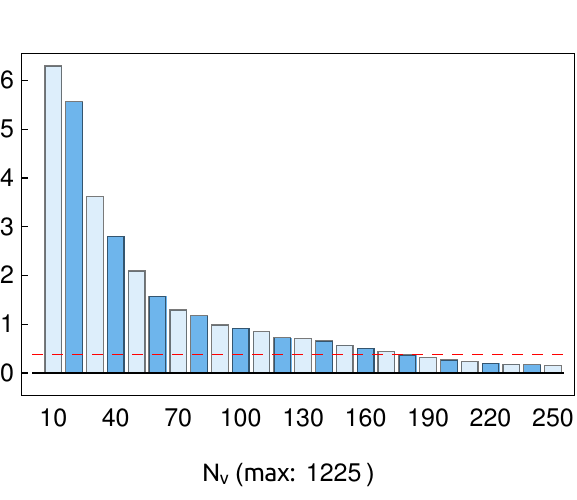} \\
\end{tabular}
 \caption{\label{enfig} Errors in the SVD-CC3 correlation energies (in mH) as a 
function of the SVD 
subspace size ($N_v$) with respect to the exact (uncompressed) CC3 method. The 
results are given 
for 
the molecules: water (first row), methane (second row), and carbon monoxide 
(third row) calculated 
by using the basis sets cc-pVTZ (first column), cc-pVQZ (second column), and 
cc-pV5Z (third 
column). 
The horizontal red dashed line marks the $1$ kJ/mol accuracy threshold (the 
chemical accuracy). The 
maximum possible size of the SVD space is given below each graph.}
\end{figure*}

Another important property of the SVD-CC3 method is the convergence pattern of 
the correlation 
energies to the exact results as a function of $N_v$. A regular and smooth 
convergence pattern is a 
highly desirable property as it allows to verify that the results are saturated 
with respect to 
$N_v$ and even estimate the error of the calculations. To address this issue we 
plot the error in 
the SVD-CC3 correlation energy as a function of $N_v$, see Fig. \ref{enfig}. We 
have selected three 
test systems: water, methane, and carbon monoxide which represent the best, an 
average, and the 
worst performance of the SVD-CC3 method in terms of the value of $\rho$ required 
to reach the 
chemical accuracy (based on Table \ref{tab1}). In general, the decay of the 
error is very regular 
and, in most cases, the convergence rate is close to exponential. This allows to 
reach the desired 
limit in a controllable fashion. The only exception from this rule are the 
results in small basis 
sets as illustrated on the left panel of Fig. \ref{enfig}. One can see that in 
small basis sets the 
SVD-CC3 method has a tendency to overshoot the correlation energy and then 
converge to the exact 
result from below. However, this behavior is not very troublesome as the scale 
of the overshooting 
is relatively minor --- around $0.1-0.2$ kJ/mol. Moreover, it is not observed in 
larger basis sets.

\begin{figure}[t]
\begin{center}
 \includegraphics[scale=0.75]{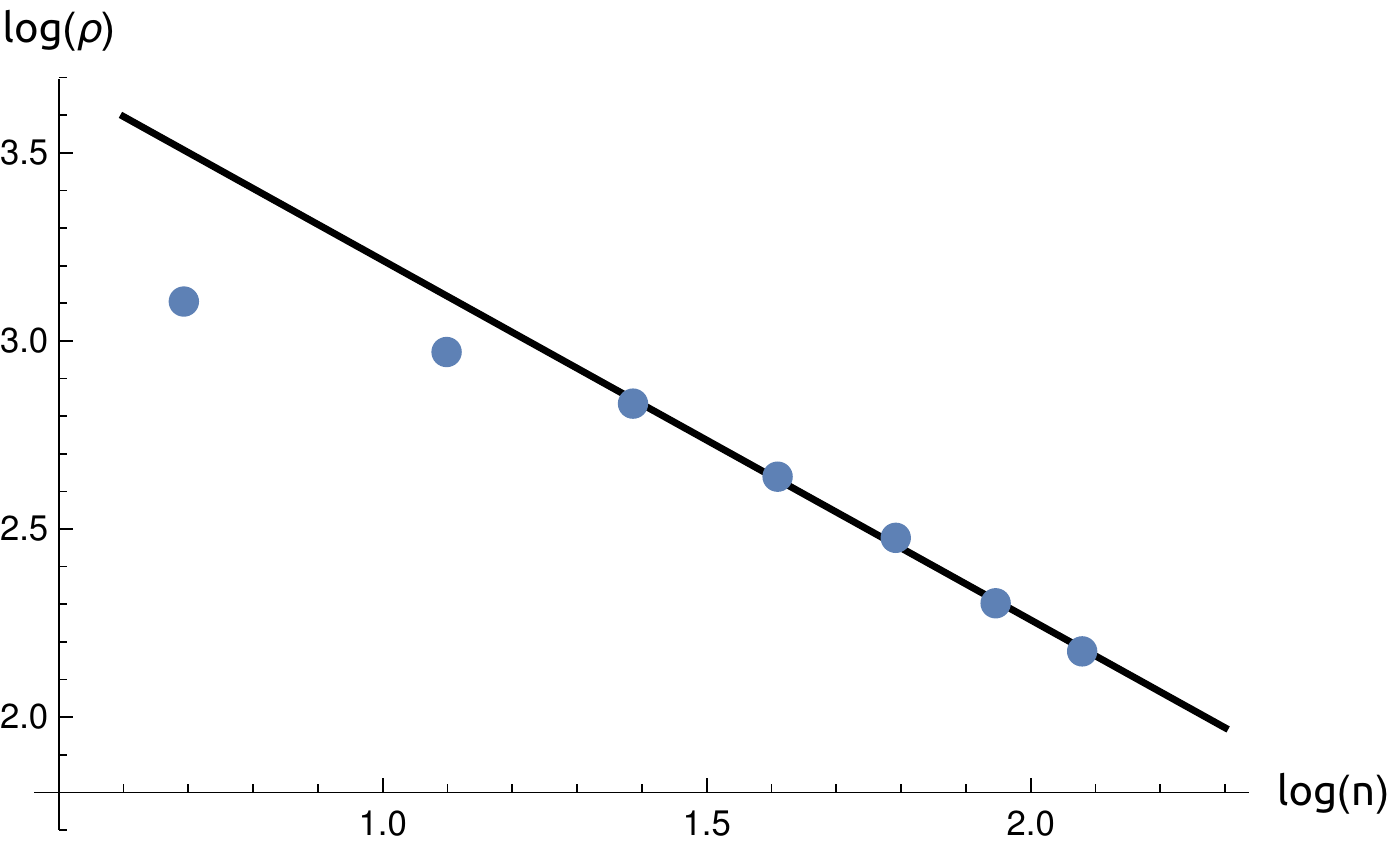} 
\end{center}
 \caption{\label{asymrho} The logarithm of the optimal value of $\rho$ (blue 
dots) for the chains of equidistant beryllium atoms, $(\mbox{Be})_n$, 
$n=2,3,\ldots,8$, as a function of the logarithm of the chain length, 
$n$. The black solid line is the least-squares linear fit to the data points 
$n=4,\ldots,8$.}
\end{figure}

While the results discussed above prove that the optimal value of $\rho$ is 
transferable between systems of a similar size, the applicability of SVD-CC 
methods to larger molecules strongly depends on the asymptotic behavior of 
$\rho$ for large $\mathcal{N}$. To investigate this issue we performed calculations for linear 
chains of equidistant beryllium atoms, $(\mbox{Be})_n$ with $n=2,3,\ldots,8$, 
employing the cc-pVDZ basis set. For each $n$ the optimal value of $\rho$ 
sufficient to reach a 
constant accuracy of $\approx1\,$kj/mol per atom was recorded. The distance between the atoms was 
set to 
$2.45\,$\AA{} --- approximately equal to the equilibrium bond length in 
$\mbox{Be}_2$ molecule \cite{patkowski07,lesiuk15}. Let us assume 
that the optimal $\rho$ behaves asymptotically as $\mathcal{N}^{-\alpha}$, 
where $\alpha$ is a parameter independent of $\mathcal{N}$. In other words, on a log-log scale the 
plot of $\rho$ as a function of
$\mathcal{N}$ (and thus the chain length, $n$, since $N\propto n$) should be a 
straight line. In 
Fig. \ref{asymrho} we plot $\log\rho$ against $\log n$ for $(\mbox{Be})_n$ with 
$n=2,3,\ldots,8$, together with the corresponding least-squares fit (the 
outliers $n=2,3$ were eliminated from the fitting procedure). The 
accuracy of the fit is surprisingly good as measured by the coefficient of 
determination, $R^2=0.998$. From the fit we also empirically 
determine the value of the parameter $\alpha\approx0.96$. This 
proves that the optimal 
value of $\rho$ decreases with the system size according to an inverse power 
law. The fact that $\alpha$ determined in this way is so close to 
the unity also suggests that asymptotically $\rho$ may be simply 
inversely proportional to $\mathcal{N}$.


In Table \ref{tab2} we present exemplary timings obtained with 
our pilot implementation in the \textsc{Gamess} package for methanol molecule 
in the cc-pVXZ basis sets, X=D,T,Q. We present total timings of the SVD-CC3 
calculations that can be compared directly with the uncompressed CC3, but also 
break down total timings into the most important components (CCSD iterations, 
Golub-Kahan bidiagonalization etc.). Three typical values of the compression 
factor, $\rho=5,\,10,\,15\%$, are considered. To allow for a fair comparison all 
of the aforementioned calculations were performed in the same computational 
environment (single-core AMD Opteron\texttrademark Processor 6174) without 
parallel execution. Moreover, the convergence thresholds and other parameters 
were kept fixed, and the DIIS convergence accelerator\cite{scuseria86} was 
turned off. The timings given in Table \ref{tab2} show that for cc-pVTZ and 
cc-pVQZ basis sets some savings with respect to the uncompressed CC3 can be 
obtained provided that $\rho<10\%$. More importantly, however, one can see that 
the cost of Golub-Kahan bidiagonalization is comparable to the cost of CCSD 
iterations. Note that the time spent on diagonalization of the $T_2$ amplitudes 
was not given in Table \ref{tab2}. This is because the cost of this procedure 
is not significant compared to the other steps of the Golub-Kahan 
bidiagonalization - for example, in the cc-pVQZ basis set the diagonalization 
of the $T_2$ amplitudes takes only about 3 minutes.

\subsection*{Relative energies}

While the results shown in the previous section prove that the compressed CC3 
method performs very 
well in recovering total correlation energies of molecular systems, the energy 
differences are of 
principal interest in most applications. Many quantum chemistry methods benefit 
from a systematic 
cancellation of errors in evaluation of energy differences. This leads to a 
significant improvement 
in their capabilities and thus it is important to determine whether the 
compressed CC methods also 
benefit from this phenomenon. To this end, we performed SVD-CC3 calculation of 
the interaction 
energies of several molecular complexes and compared the results with the 
uncompressed CC3 method. 
The test examples include two hydrogen-bonded complexes (H$_2$O dimer, HF 
dimer), a system bound 
by dispersion forces (CH$_4-$BH$_3$) and a molecular complex of a mixed 
induction-dispersion 
character (CH$_4-$HF). A rather diverse set of model systems were selected to 
avoid bias towards any 
particular type of interaction. The geometries were taken from the A24 test set 
of 
\v{R}ez\'{a}\u{c} 
and Hobza \cite{rezac13a}. The aug-cc-pVTZ basis set \cite{dunning92} was used 
in all calculations 
presented in this section.

Interaction energies calculated for the aforementioned systems are given in 
Table \ref{tab3}. We 
list the data starting with $N_v=20$ because smaller SVD spaces typically give 
results which are by 
an order of magnitude wrong and thus of little practical use. Results given in 
Table \ref{tab3} 
lead 
to two important conclusions. First, there is a systematic error cancellation in 
evaluating the 
energy differences with compressed CC methods. For example, the error of the 
SVD-CC3 interaction 
energy for the HF dimer stabilizes below 1 kJ/mol at the compression levels of 
around $\rho=5\%$. 
To 
reach the same level of accuracy in total energies of the HF dimer and of the 
separated monomers 
one 
needs $\rho=10.2\%$ and $\rho=10.9\%$, respectively. Therefore, a significant 
fraction of the error 
in raw energies canceled out in evaluation of $E_{\mbox{\scriptsize int}}$. 
This allows for 
reliable estimation of the interaction energies with compression factors as 
small as $5\%$.

\begin{figure}[t]
\begin{center}
 \includegraphics[scale=1.10]{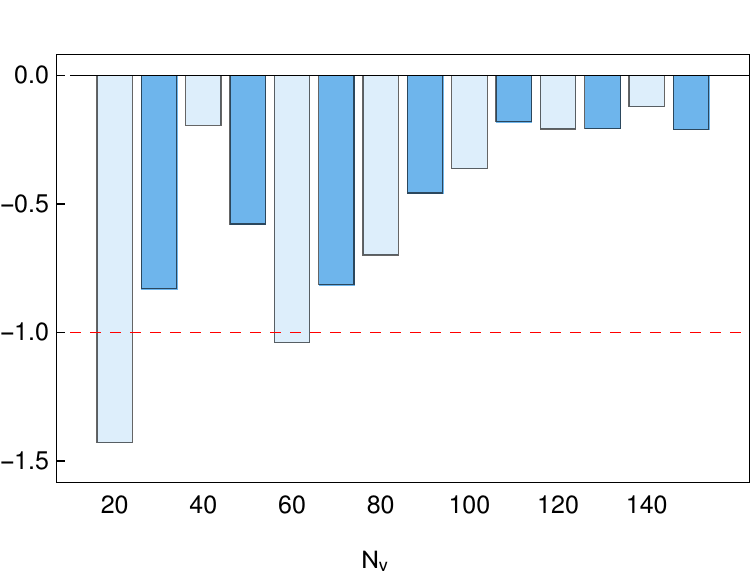} 
\end{center}
 \caption{\label{hfdimfig} Errors in the interaction energy of the HF dimer (in 
kJ/mol) calculated 
with the SVD-CC3 method as a function of the SVD subspace size ($N_v$). The 
horizontal red dashed 
line marks the $1$ kJ/mol accuracy threshold (the chemical accuracy). The 
maximum possible size of 
the SVD space is 1280.}
\end{figure}

The second conclusion from Table \ref{tab3} is that the convergence of the 
SVD-CC3 results towards 
the exact value as a function of $N_v$ is not a smooth as in the case of total 
energies discussed 
in 
the previous section. This is illustrated in Fig. \ref{hfdimfig} for the  HF 
dimer. While the 
general convergence pattern is still clear, some accidental cancellations cause 
the error for 
several $N_v$ to be significantly smaller than the overall trend in the data 
would suggest. 
Additionally, the results given in Table \ref{tab3} suggest that the natural 
noise level of the 
SVD-CC3 method is $0.1-0.2$ kJ/mol in the current implementation. This is to be 
expected due to, 
\emph{e.g.} rather loose values of some thresholds set in our program. 
Nonetheless, this noise 
level 
is significantly below the accuracy of the uncompressed CC3 method itself.

As the second illustration of the cancellation of errors in the SVD-CC3 method 
we consider energy 
differences between different geometries of the same molecule. The internal 
rotation of the 
hydrogen 
atom around the C$-$O bond in the formic acid is used as a model system. The 
formic acid has two 
stable conformers (trans and cis) characterized by H$-$C$-$O$-$H dihedral angle 
of 
$0^{\mbox{\scriptsize o}}$ and $180^{\mbox{\scriptsize o}}$, respectively. We 
calculated energy 
differences between the trans and cis conformers, height and location of the 
corresponding 
rotational barrier. This is achieved by performing potential energy scan varying 
the H$-$C$-$O$-$H 
dihedral angle in steps of $15^{\mbox{\scriptsize o}}$ and keeping the remaining 
internal 
coordinates of the molecule fixed. The results obtained with SVD-CC3 method are 
shown in Table 
\ref{tab4} and compared with the uncompressed CC3 method. Even with the 
compression factor as small 
as $\rho\approx1\%$ the SVD-CC3 results are wrong only by about $1-2$ kJ/mol. 
For all intents and 
purposes the values obtained with $\rho\approx4\%$ are essentially 
indistinguishable from the 
uncompressed CC3 results. This suggest that the cancellation of errors between 
different geometries 
of the same molecule is even more substantial than in the calculation of 
interaction energies. We 
expect the same conclusion to be valid also in other processes which do not 
involve breaking of 
chemical bonds and other drastic rearrangements of electronic densities.

\subsection*{Bond-breaking processes}

Let us point out that for a majority of molecules which have been considered 
thus far in the paper 
the CCSD(T) method gives acceptable results. In other words, $T_3^{[2]}$ is a 
good approximation to 
the exact $T_3$ cluster operator for these systems and it is reasonable to use 
it as a source of 
amplitudes for SVD. However, one can argue that the performance of the SVD-CC 
methods will be much 
worse outside the regime of applicability of the $T_3^{[2]}$ approximation, e.g. 
in cases where the 
CCSD(T) method fails due to significant static correlation effects. Fortunately, 
the results given 
in Ref. \onlinecite{hino04} suggest that this is not true and in this section we 
provide an 
additional verification of this claim.

One prominent example of a process where the CCSD(T) method fails both 
quantitatively and 
qualitatively is breaking of a chemical bond 
\cite{chan58,laidig87,piecuch02,piecuch04}. It is 
known 
that CCSD(T) gives nonphysical results even for a relatively simple cleavage of 
a single bond. To 
investigate the performance of the SVD-CC3 method in description of this process 
we performed 
calculations for bond-breaking reactions in two molecules (F$_2$ and CH$_4$). We 
stretched the 
F$-$F 
bond and one of the C$-$H bonds from $0.75\cdot R_e$ to $3.00\cdot R_e$, where 
$R_e$ is the 
equilibrium bond length, and calculated the total energies of the system with 
the following 
methods: 
\begin{itemize}
 \item conventional CCSD and CCSD(T);
 \item uncompressed CC3;
 \item completely renormalized CR-CCSD(T), CR-CC(2,3) methods of Piecuch 
\emph{et al.} 
\cite{piecuch02b};
 \item SVD-CC3 method with $\rho\approx5,10,\ldots,25\%$;
 \item full CCSDT method.
\end{itemize}
The last method gives negligible errors with respect to FCI for both systems and 
thus can be used 
as 
a reference. The aug-cc-pVTZ basis set is employed in all calculations reported 
in this section; 
this is the largest basis set we could use with the CCSDT method. The reference 
calculations were 
performed with the \textsc{AcesII} program package \cite{aces2}.

The results of the calculations are given in Table \ref{tab5} for the F$_2$ 
molecule and in Table 
\ref{tab6} for the CH$_4$ molecule. The first observation is that the standard 
CCSD(T) method 
behaves unexpectedly well near the bottom of the potential energy curve. The 
errors of 
CCSD(T) for $R=R_e$ are one of the smallest among the considered methods, but 
the accuracy of 
CCSD(T) deteriorates rapidly when the bond is stretched. As quickly as for 
$1.25\cdot R_e$ SVD-CC3 
outperforms CCSD(T). Moreover, for $R> 2.0\cdot R_e$ the error of CCSD(T) 
becomes catastrophic 
while SVD-CC3 retains a fairly constant level of accuracy. Additionally, the 
SVD-CC3 method with 
$\rho\approx10,\,15\%$ offers a considerable improvement over CR-CCSD(T) for 
nearly all internuclear distances. 
This behavior is somewhat more pronounced for the F$_2$ molecule than for 
CH$_4$. The performance of CR-CC(2,3) and SVD-CC3 is quite similar -- both 
methods give nearly equivalent results for CH$_4$, but SVD-CC3 performs somewhat 
better for F$_2$. Note that somewhat larger $\rho$ are required here to maintain a satisfactory 
level of accuracy compared with some of the previous examples given in the paper. However, the 
$\rho\approx15\%$ is sufficient for most purposes and the results with $\rho\approx20\%$ differ 
insignificantly from the uncompressed CC3 values. To sum up, SVD-CC3 method is able to describe the 
single-bond 
breaking process with a 
consistent accuracy of a few kJ/mol, offering a dramatic improvement over 
perturbative methods such as CCSD(T) and CR-CCSD(T), and without sacrificing 
much accuracy for molecules in near-equilibrium 
geometry.

We have to mention that the comparison between the CR-CCSD(T) (and related 
methods) and SVD-CC3 is 
not entirely fair because the latter method is iterative and thus inherently 
more expensive. In the 
best case scenario, SVD-CC3 is expected to be twice as expensive as CR-CCSD(T). 
In practice, we 
found this ratio to be about 3$-$4 with the compression rates considered here. 
This overhead is 
substantial but still acceptable. By comparison, CCSDT computations are hundreds 
of times more 
expensive. We could not compare the SVD-CC3 results with 
$\Lambda$CCSD(T) methods
\cite{kucharski98b} as it is not implemented in any quantum chemistry program 
available to us.

The results provided in this section show that $T_3^{[2]}$ can be used as an 
initial source of approximate 
triple excitation amplitudes for SVD even in a non-perturbative regime. It 
appears that the 
perturbative expressions manage to correctly identify the important excitations 
and their rough 
relative importance but they overestimate the overall effect of the triple 
excitations. 
Fortunately, 
this does not prevent the SVD procedure from extracting the most important 
information about the 
exact $T_3$ amplitudes.

\section*{Summary and conclusions}

In this work we have presented a novel method for calculating SVD of the CC 
triple excitation 
amplitudes tensor. Our technique is based on the Golub-Kahan bidiagonalization 
and does not require 
$T_3$ to be stored. Moreover, the cost of the procedure is relatively small - 
comparable to several 
CCSD iterations. We have illustrated the usefulness of the new method by 
computing SVD of an 
approximate (perturbative) triple excitations amplitudes tensor, and 
subsequently using the most 
significant singular vectors as a basis for expansion of the $T_3$ cluster 
operator in the 
iterative 
CC3 method.

The resulting SVD-CC3 method has been tested by calculating total correlation 
energies for a set of 
small and medium-sized molecules and comparing the results with the exact 
(uncompressed) CC3 
method. 
We have shown that the compression factors in the range $\rho=5-15\%$ are 
sufficient to reach the 
chemical accuracy ($\approx 1$ kJ/mol) of the results. Moreover, we have shown 
that SVD-CC3 method benefits from a 
substantial error cancellation in evaluation of, \emph{e.g.} interaction 
energies or energy 
differences between various geometries of the same molecule. It has been found 
that the compression 
factors of about $\rho\approx 5\%$ are practically sufficient in evaluation of 
relative energies.

Finally, we have investigated the performance of the SVD-CC3 method in the 
processes of breaking of 
a single bond. The results indicate that SVD-CC3 is free of instabilities found 
in CCSD(T) and 
other 
perturbative methods, and is able to describe single bond-breaking with an 
accuracy of a few 
kJ/mol. 
Let us also point out that SVD-CC3 is a black-box method in the sense that only 
one input parameter 
($\rho$) must be supplied by the user. The remaining internal thresholds and 
parameters have been 
kept constant during all calculations reported in this work and no significant 
technical 
difficulties have been observed. In particular, there is no need to specify any 
active orbital 
space 
which is both tedious and requires a considerable physical insight into the 
system being studied.

Since SVD of an approximate triple excitation amplitudes tensor can now be 
obtained relatively cheaply, \emph{i.e.} with a cost much lower than 
$\mathcal{N}^8$, the idea of the full SVD-CCSDT method becomes viable. This 
requires careful factorization of all terms in the CC triples residual, and 
possibly application of additional decomposition schemes for the $t_{XYZ}$ 
tensor. However, one can expect the compressed SVD-CCSDT to possess an accuracy 
level comparable to the full CCSDT, but at a significantly reduced cost.
This would provide a 
relatively inexpensive computational method capable of handling single 
bond-breaking, biradical 
species, etc., maintaining the chemical accuracy and the black-box character of 
single-reference CC 
methods. To be able to describe double bond-breaking reliably one requires also 
quadruple 
excitations to be included in the theoretical model. To this end, various 
perturbative quadruple 
corrections \cite{kucharski89,bartlett90,kucharski98,kucharski01} calculated on 
top of SVD-CCSDT 
seem attractive, extending  capabilities of the SVD-CC family of methods even 
further.

Finally, let us point out that in the present work we have avoided using any 
approximations to the 
CC equations other than the SVD itself. However, it has been shown that 
techniques such as density 
fitting \cite{whitten73,dunlap79,vahtras93,feyer93,rendell94} or Cholesky 
decomposition 
\cite{beebe77,roeggen86,koch03,auilante07} are able to reduce the storage and 
computational 
requirements of the CC methods without a significant loss in the accuracy 
\cite{deprince13}. 
Therefore, it would be reasonable to incorporate this techniques in future 
SVD-CC implementations along with an efficient 
parallelization. \cite{prochnow10}
 
\subsection*{Acknowledgements}

I would like to thank Prof. G. Cha\l asi\'{n}ski and Prof. B. Jeziorski for 
reading and commenting on the manuscript, and Dr. T. Korona for fruitful 
discussions concerning the efficiency of coupled-cluster programs. The author is 
indebted to Prof. P. Piecuch for finding a numerical error in an early version 
of the manuscript, and to J. Emiliano Deustua, I. Magoulas, and S. H. Yuwono 
for sharing results of their coupled-cluster calculations which helped to 
eliminate the problem. This work was supported by the National Science Center, 
Poland within the project 2017/27/B/ST4/02739.

\clearpage

\bibliography{ref}

\clearpage

\clearpage

\clearpage

\clearpage

\begin{table}[t]
\caption{Results of the SVD-CC3 calculations of the total correlation energy. 
For each basis set 
the 
size of the SVD subspace ($N_v$) and the corresponding compression factor 
($\rho$) are reported 
which are sufficient to reach the chemical accuracy of the results 
($1\,\mbox{kJ/mol} \approx 
0.4\,\mbox{mH}$). The values of $\rho$ are given in percents.}
\label{tab1}
\begin{center}
 \begin{tabular}{lrrrrrrrr}
\hline\hline
& \multicolumn{2}{c}{cc-pVDZ} 
 & \multicolumn{2}{c}{cc-pVTZ} 
 & \multicolumn{2}{c}{cc-pVQZ} 
 & \multicolumn{2}{c}{cc-pV5Z} \\
 & \multicolumn{1}{c}{$N_v$} & \multicolumn{1}{c}{$\rho$} 
 & \multicolumn{1}{c}{$N_v$} & \multicolumn{1}{c}{$\rho$} 
 & \multicolumn{1}{c}{$N_v$} & \multicolumn{1}{c}{$\rho$} 
 & \multicolumn{1}{c}{$N_v$} & \multicolumn{1}{c}{$\rho$} \\
\hline
H$_2$O     & 13 & 13.0 & 23  & 8.5  & 37  & 6.7  & 63  & 6.4  \\
HCN        & 13 & 7.1  & 47  & 9.9  & 97  & 10.4 & 188 & 11.7 \\
CH$_4$     & 24 & 16.2 & 40  & 9.8  & 78  & 9.2  & 117 & 7.6  \\
H$_2$CO    & 32 & 13.4 & 61  & 9.5  & 151 & 11.7 & --- & ---  \\
C$_2$H$_4$ & 43 & 13.2 & 115 & 13.3 & 219 & 12.3 & --- & ---  \\
NH$_3$     & 18 & 14.9 & 29  & 8.7  & 54  & 7.7  & 97  & 7.7  \\
H$_2$S     & 34 & 15.8 & 52  & 9.9  & 66  & 6.4  & 210 & 11.6 \\
C$_2$H$_2$ & 18 & 8.2  & 45  & 7.9  & 106 & 9.3  & --- & ---  \\
CO         & 19 & 13.0 & 43  & 11.6 & 88  & 12.2 & 178 & 14.5 \\
CH$_3$OH   & 47 & 13.2 & 83  & 8.5  & 200 & 10.0 & --- & ---  \\
N$_2$H$_4$ & 49 & 13.7 & 78  & 8.1  & 240 & 12.1 & --- & ---  \\
H$_2$O$_2$ & 36 & 13.6 & 62  & 8.6  & 179 & 12.4 & --- & ---  \\
BF$_3$     & 61 & 9.4  & 82  & 4.9  & 232 & 7.1  & --- & ---  \\
NCCN       & 59 & 10.6 & 181 & 13.0 & 299 & 11.1 & --- & ---  \\
LiF        & 17 & 12.7 & 29  & 8.9  & 41  & 6.5  & 54  & 5.1  \\
\hline
average    & --- & 12.5 & --- & 9.4 & --- & 9.7 & --- & 9.2 \\
std. dev.  & --- & 2.6  & --- & 2.1 & --- & 2.3 & --- & 3.4 \\
\hline\hline
\end{tabular} 
\end{center}
\end{table} 

\clearpage

\begin{table}[t]
\caption{Single-core timings of the SVD-CC3 method for the methanol molecule in 
the cc-pVXZ basis sets, X=D,T,Q. The values ''total SVD-CC3`` include 
timings for all steps of the SVD-CC3 calculations described in the text 
(CCSD, Golub-Kahan etc.) and can be directly compared with the uncompressed CC3 
method (the last row).}
\label{tab2}
\begin{center}
\begin{tabular}{c|rrr}
\hline\hline
 task & \multicolumn{3}{c}{timings (min)} \\
      & cc-pVDZ & cc-pVTZ & cc-pVQZ \\
\hline
Hartree-Fock & 0.0 & 0.7 & 13.3 \\
four-index trans. (without VVVV) & 0.1 & 0.7 & 10.9 \\
CCSD (no DIIS) & 1.4 & 30.0 & 600.0 \\
\hline
Golub-Kahan bidiag. ($\rho\approx 5\%$) & 0.9 & 29.8 & 245.9 \\
CC iterations ($\rho\approx 5\%$) & 1.7 & 33.4 & 727.4 \\
total SVD-CC3 ($\rho\approx 5\%$) & 4.1 & 94.6 & 1597.5 \\
\hline
Golub-Kahan bidiag. ($\rho\approx 10\%$) & 1.2 & 34.1 & 318.4 \\
CC iterations ($\rho\approx 10\%$) & 2.2 & 44.4 & 765.2 \\
total SVD-CC3 ($\rho\approx 10\%$) & 4.9 & 109.9 & 1707.8 \\
\hline
Golub-Kahan bidiag. ($\rho\approx 15\%$) & 1.5 & 58.0 & 437.3 \\
CC iterations ($\rho\approx 15\%$) & 2.9 & 66.1 & 845.1 \\
total SVD-CC3 ($\rho\approx 15\%$) & 5.9 & 155.5 & 1906.6 \\
\hline
total uncompressed CC3 & 4.1 & 131.8 & 1704.3 \\
\hline\hline
\end{tabular}  
\end{center}
\end{table}

\clearpage

\begin{table}[t]
\caption{SVD-CC3 interaction energies ($E_{\mbox{\scriptsize int}}$) of several 
molecular complexes 
and the corresponding absolute errors ($\delta E_{\mbox{\scriptsize int}}$) with 
respect to the 
uncompressed CC3 method as a function of the number of singular vectors included 
in the expansion, 
$N_v$. All results were obtained with the aug-cc-pVTZ basis set. The reference 
CC3 interaction 
energies are given in the last line. The values of $\rho$ are given in percents 
and the interaction 
energies in kJ/mol.}
\label{tab3}
\begin{center}
 \begin{tabular}{rrrrrrrrrrrrrrrr}
\hline\hline
$N_v$ & \multicolumn{3}{c}{H$_2$O$-$H$_2$O} 
      & \multicolumn{3}{c}{HF$-$HF} 
      & \multicolumn{3}{c}{CH$_4-$HF} 
      & \multicolumn{3}{c}{CH$_4-$BH$_3$} \\
      & \multicolumn{1}{c}{$\rho$} 
      & \multicolumn{1}{c}{$E_{\mbox{\scriptsize int}}$} 
      & \multicolumn{1}{c}{$\delta E_{\mbox{\scriptsize int}}$}
      & \multicolumn{1}{c}{$\rho$} 
      & \multicolumn{1}{c}{$E_{\mbox{\scriptsize int}}$} 
      & \multicolumn{1}{c}{$\delta E_{\mbox{\scriptsize int}}$}
      & \multicolumn{1}{c}{$\rho$} 
      & \multicolumn{1}{c}{$E_{\mbox{\scriptsize int}}$} 
      & \multicolumn{1}{c}{$\delta E_{\mbox{\scriptsize int}}$}
      & \multicolumn{1}{c}{$\rho$} 
      & \multicolumn{1}{c}{$E_{\mbox{\scriptsize int}}$} 
      & \multicolumn{1}{c}{$\delta E_{\mbox{\scriptsize int}}$} \\
\hline
 20 & 1.2 & 28.2 & 4.3 & 1.6  & 20.4 & 1.4 & 1.0 & 8.6 & 1.1 & 0.5 & 7.5 & 0.3 
\\
 30 & 1.7 & 31.4 & 7.4 & 2.3  & 21.0 & 0.8 & 1.5 & 8.7 & 1.0 & 0.9 & 6.1 & 1.7 
\\
 40 & 2.3 & 23.9 & 0.0 & 3.1  & 21.6 & 0.2 & 2.0 & 9.6 & 0.1 & 1.4 & 6.8 & 1.0 
\\
 50 & 2.9 & 24.9 & 0.9 & 3.9  & 21.3 & 0.5 & 2.5 & 9.6 & 0.1 & 1.8 & 6.9 & 0.9 
\\
 60 & 3.5 & 25.3 & 1.3 & 4.7  & 20.8 & 1.0 & 3.0 & 9.3 & 0.4 & 2.3 & 7.0 & 0.8 
\\
 70 & 4.0 & 24.3 & 0.3 & 5.5  & 21.0 & 0.8 & 3.6 & 9.1 & 0.6 & 2.7 & 7.2 & 0.6 
\\
 80 & 4.6 & 23.9 & 0.1 & 6.3  & 21.1 & 0.7 & 4.1 & 9.1 & 0.6 & 3.2 & 7.3 & 0.5 
\\
 90 & 5.2 & 23.8 & 0.2 & 7.0  & 21.4 & 0.4 & 4.6 & 9.3 & 0.4 & 3.6 & 7.4 & 0.4 
\\
100 & 5.8 & 23.8 & 0.2 & 7.8  & 21.5 & 0.3 & 5.1 & 9.3 & 0.4 & 4.1 & 7.5 & 0.3 
\\
110 & 6.3 & 23.9 & 0.1 & 8.6  & 21.6 & 0.2 & 5.6 & 9.2 & 0.5 & 4.6 & 7.6 & 0.2 
\\
120 & 6.9 & 23.9 & 0.1 & 9.4  & 21.6 & 0.2 & 6.1 & 9.4 & 0.3 & 5.0 & 7.6 & 0.2 
\\
130 & 7.5 & 23.9 & 0.1 & 10.2 & 21.6 & 0.2 & 6.6 & 9.5 & 0.2 & 5.9 & 7.6 & 0.2 
\\
140 & 8.1 & 23.9 & 0.1 & 10.9 & 21.7 & 0.2 & 7.1 & 9.5 & 0.2 & 6.4 & 7.7 & 0.1 
\\
150 & 8.6 & 23.9 & 0.1 & 11.7 & 21.7 & 0.1 & 7.6 & 9.5 & 0.2 & 6.8 & 7.7 & 0.1 
\\
\hline
max & --- & 24.0 & --- & --- & 21.8 & --- & --- & 9.7 & --- &  --- & 7.8 & --- & 
\\
\hline\hline
\end{tabular} 
\end{center}
\end{table}

\clearpage

\begin{table}[t]
\caption{SVD-CC3 results for the internal rotation of the hydrogen atom around 
the H$-$C$-$O$-$H 
dihedral angle in the formic acid;
$\Delta E_{\mbox{\scriptsize cis/trans}}$ is the energy difference between the 
cis and trans 
conformers, $\theta_{\mbox{\scriptsize max}}$ is the value of the dihedral angle 
corresponding to 
the height of the barrier, and $\Delta E_{\mbox{\scriptsize barrier}}$ is the 
energy difference 
between the trans conformer and the barrier maximum. The energies are given in 
kJ/mol and the 
angles 
are given in degrees.}
\label{tab4}
\begin{center}
\begin{tabular}{ccccccccc}
\hline\hline
 method & $\Delta E_{\mbox{\scriptsize cis/trans}}$ & $\theta_{\mbox{\scriptsize 
max}}$ & $\Delta 
E_{\mbox{\scriptsize barrier}}$ \\
 \hline
 SVD-CC3, $\rho\approx1\%$ & 19.3 & 92.7 & 57.5 \\
 SVD-CC3, $\rho\approx2\%$ & 20.5 & 93.0 & 59.2 \\
 SVD-CC3, $\rho\approx3\%$ & 20.8 & 92.7 & 59.7 \\
 SVD-CC3, $\rho\approx4\%$ & 20.3 & 92.7 & 59.6 \\
 \hline
 uncompressed CC3 & 20.3 & 92.9 & 59.5 \\
 \hline\hline
\end{tabular}  
\end{center}
\end{table}

\clearpage

\begin{landscape}
\begin{table}
\caption{A comparison of total energies of the F$_2$ molecule calculated with 
various CC methods at 
several internuclear distances (aug-cc-pVTZ basis set). The first row shows 
total CCSDT energies 
(in the atomic units); absolute errors with respect to the CCSDT values (in mH) 
are given in the 
remaining rows. The equilibrium bond length of F$_2$ is $R_e=1.27455\,$\AA{}. 
The $1s$ core 
orbitals of the fluorine atoms were kept frozen in all correlated calculations.}
\label{tab5}
\begin{center}
\begin{tabular}{lrrrrrrrrrr}
\hline\hline
 method & $0.75\cdot R_e$ & \multicolumn{1}{c}{\;\;\;$R_e$} & $1.25\cdot R_e$ & 
$1.50\cdot R_e$ & 
$2.00\cdot R_e$ & $3.00\cdot R_e$ \\
 \hline
 CCSDT      & $-198.928\,781$ & $-199.297\,867$ & $-199.302\,636$ & 
$-199.272\,578$ & 
$-199.253\,853$ & $-199.253\,283$ \\
 \hline
 CCSD       & 11.598 & 16.697 & 25.421   & 38.339   & 60.492    & 70.387 \\
 CCSD(T)    & 0.077  & 0.065  & $-$0.268 & $-$2.510 & $-$19.214 & $-$41.951 \\
 CR-CCSD(T) & 1.648  & 2.759  & 5.060    & 8.947    & 14.414    & 14.030 \\
 CR-CC(2,3) & 0.008  & 0.218  & 0.927    & 2.741    & 5.504     & 5.216 \\
 CC3        & $-$0.444 & $-$0.623 & $-$0.695 & $-$0.701 & $-$1.855 & $-$4.212 \\
 SVD-CC3, $\rho\approx5\%$  & 1.289 & 1.770 & 2.928 & 4.711 & 7.033 & 6.878 \\
 SVD-CC3, $\rho\approx10\%$ & $-$0.326 & $-$0.357 & 0.274 & 0.772 & 1.159 & $-$0.523 \\
 SVD-CC3, $\rho\approx15\%$ & $-$0.613 & $-$0.636 & $-$0.349 & 0.102 & $-$0.758 & $-$2.830 \\
 SVD-CC3, $\rho\approx20\%$ & $-$0.567 & $-$0.697 & $-$0.601 & $-$0.428 & $-$1.439 & $-$3.514 \\
 SVD-CC3, $\rho\approx25\%$ & $-$0.537 & $-$0.681 & $-$0.733 & $-$0.696 & $-$1.732 & $-$4.023 \\
 \hline\hline
\end{tabular}  
\end{center}
\end{table}
\end{landscape}

\clearpage

\begin{landscape}
\begin{table}
\caption{A comparison of total energies of the CH$_4$ molecule calculated with 
various CC methods 
at several internuclear distances (aug-cc-pVTZ basis set). The first row shows 
total CCSDT energies 
(in the atomic units); absolute errors with respect to the CCSDT values (in mH) 
are given in the 
remaining rows. The equilibrium C$-$H bond length is $R_e=1.0870\,$\AA{}. The 
$1s$ core orbital of 
the carbon atom was kept frozen in all correlated calculations.}
\label{tab6}
\begin{center}
\begin{tabular}{lrrrrrrrrrr}
\hline\hline
 method & $0.75\cdot R_e$ & $R_e$ & $1.25\cdot R_e$ & $1.50\cdot R_e$ & 
$2.00\cdot R_e$ & 
$3.00\cdot 
R_e$ \\
 \hline
 CCSDT      & $-40.358\,055$ & $-40.441\,301$ & $-40.413\,646$ & $-40.367\,481$ 
& $-40.295\,998$ & 
$-40.254\,907$ \\
 \hline
 CCSD       & 6.498 & 6.932 & 7.570 & 8.575 & 12.652 & 23.892 \\
 CCSD(T)    & 0.372 & 0.392 & 0.437 & 0.512 & 0.298  & $-$14.054 \\
 CR-CCSD(T) & 1.114 & 1.223 & 1.411 & 1.744 & 3.149  & 5.427 \\
 CR-CC(2,3) & 0.318 & 0.308 & 0.339 & 0.432 & 0.934  & 0.475 \\
 CC3        & 0.190 & 0.173 & 0.171 & 0.203 & 0.392  & 0.146 \\
 SVD-CC3, $\rho\approx5\%$  & 1.437 & 1.476 & 1.621 & 1.932 & 2.399 & 3.651 \\
 SVD-CC3, $\rho\approx10\%$ & 0.338 & 0.311 & 0.360 & 0.497 & 1.107 & 1.819 \\
 SVD-CC3, $\rho\approx15\%$ & 0.265 & 0.251 & 0.253 & 0.348 & 0.612 & 0.643 \\
 SVD-CC3, $\rho\approx20\%$ & 0.227 & 0.200 & 0.199 & 0.242 & 0.514 & 0.448 \\
 SVD-CC3, $\rho\approx25\%$ & 0.203 & 0.190 & 0.183 & 0.218 & 0.450 & 0.306 \\
 \hline\hline
\end{tabular}  
\end{center}
\end{table}
\end{landscape}

\end{document}